\documentclass[12pt, a4paper]{article}
\usepackage{amsfonts}
\usepackage{amsmath}
\usepackage{amssymb}
\usepackage{amsthm}
\usepackage[toc]{appendix}
\usepackage[hiresbb]{graphicx}
\usepackage{caption}
\usepackage{subcaption}
\usepackage{booktabs}
\usepackage[utf8]{inputenc}
\usepackage{listings}
\usepackage{here}
\usepackage{float}
\usepackage{ascmac}
\usepackage{tikz}
\usepackage{url}
\usepackage{lipsum}
\usepackage{authblk}
\usepackage{eqnarray}
\usepackage{siunitx}
\usepackage[sectionbib,round]{natbib}
\newtheorem{theorem}{Theorem}[section]

\newtheorem{definition}{Definition}[section]
\newtheorem{corollary}{Corollary}[section]
\newtheorem{lemma}{Lemma}[section]
\DeclareMathOperator{\tr}{Tr}

\newcommand{\be}{\begin{equation}}
\newcommand{\ee}{\end{equation}}
\newcommand{\beq}{\begin{eqnarray*}}
\newcommand{\eeq}{\end{eqnarray*}}
\def\sym#1{\ifmmode^{#1}\else\(^{#1}\)\fi}

\title{\large{\bf{Methods of Stochastic Field Theory in Non-Equilibrium Systems \\
--- Spontaneous Symmetry Breaking of Ergodicity}}}
\author{\large{\bf{Tatsuru Kikuchi}}}
\affil{\small{\it{Faculty of Economics, The University of Tokyo,}}\\
{\it{7-3-1 Hongo, Bunkyo-ku, Tokyo 113-0033 Japan}}}
\date{\small{(\today)}}
\begin{document}
\maketitle
\begin{abstract}
Recently, a couple of investigations related to symmetry breaking phenomena, 'spontaneous stochasticity' and 'ergodicity breaking' have led to significant impacts in a variety of fields related to the stochastic processes such as economics and finance. We investigate on the origins and effects of those original symmetries in the action from the mathematical and the effective field theory points of view. It is naturally expected that whenever the system respects any symmetry, it would be spontaneously broken once the system falls into a vacuum state which minimizes an effective action of the dynamical system. 
\end{abstract}
\newpage
\section{Introduction}
Recently, a couple of investigations related to symmetry breaking phenomena, 'spontaneous stochasticity' (in the original terminology) and 'ergodicity breaking', have led to significant impacts in a variety of fields related to the stochastic processes and the others. The former one has been investigated by \citet{Mai2016}, and the latter one has investigated by \citet{Pet2013} in the context of geometric Brownian motions. There are several studies on the ergodicity breaking in a variety of fields. for instance, \citet{Tou2023} has studied criticality of the employment and personal-income dynamics, \citet{Med2021} investigated the time optimal choice of utility functions under the assumption that ergodic theories of decision-making reveal how individuals should tolerate risk in different environments, \citet{Man2023} has studied biomarker discovery for heart disease, \citet{Sac2023} has studied ergodicity transitions in condensed matter physics, and the others. \citet{Pet2019} emphasizes the importance of problem settings where the ergodicity breaking may or may not cause some practical influences, especially in the study of economics. 

At first, let us run over the important points of discussions that in dispute and need to be settled. The notion of ergodicity has been arisen in classical statistical mechanics in Physics, and it is only one of the macroscopic properties in equilibrium systems with finite size of volume. In fact, all the observables have several constraints, for instance, we are restricted in a finite time system, hence the notion of ergodicity should be regarded as an approximate nature that has been appeared in macroscopic systems. Moreover, when we seek to understand the microscopic phenomena, in which the size of the system in concern becomes much smaller and interaction for each of the particles becomes relevant. In such a situation, the notion of ergodicity cannot be applied. It is a subject of measure theory to understand the nature of $G$-invariant measures when we consider a group action to the dynamical systems. 

We investigate on a various phenomena using the methods of stochastic field theory (or the effective field theory in quantum field theory applied to the stochastic dynamics) with the use of path integral formulation. In general, we consider the case when there is a gauge symmetry $G$ which is spontaneously broken down to $H$, then it appears a number of massless Nambu-Goldstone (NG) modes. The order parameter is given by $g = g \cdot H \in G/H$. Furthermore, the quantum field theory has succeeded for more than a century in understanding a wide range of phenomena for a physics at a short distance (high-energy) to the one at a long distance (low-energy). It has been applied to a wide range of phenomena, for instance, critical dynamics, turbulence, and stochastic dynamics. The interplay between the physics at short distance and the one at large distance has also been well established via the Renormalization Group. It is also true for the statistical mechanics in which the system is weekly correlated, composed of a large enough degrees of freedom (typically, $N=10^{23}$). In probability, weekly correlated system can be regarded as a collection of random variables, and statistical properties may be originated from the central limit theorem. 

However, some complications arise when the system becomes strongly correlated, for instance, it happens in the vicinity of second-order phase transition near the critical points. In such a case, the system becomes strongly correlated at larger and larger distances, where the mean field theory fails to explain the critical phenomena. The Renormalization Group gives an appropriate way to understand the strongly correlated systems. Since the theoretical breakthrough, the method of the Wilsonian Renormalization Group makes a primary concepts in the effective field theory. In the method of the Wilsonian Renormalization Group, coarse graining variables by integrating out the high-energy (short distance) degrees of freedom is used to construct the theory at low-energy (log distance) in strongly correlated systems. For the illustrative purpose, let us show the way to construct a low-energy effective action $S_{\mathrm{eff}}$ by integrating out the high energy degrees of freedom
\be
\exp \left(-S_{\mathrm{eff}}[\phi]|_{\mu < \Lambda} \right) = \int [{\cal{D}} \phi]|_{\mu > \Lambda} \exp \left(-S[\phi]|_{\mu > \Lambda} \right) \;.
\ee
The scale invariance and self-similarity is a class of universality, which appear in critical phenomena near the fixed points. Even though it one of the practically useful tools, the methods of Renormalization Group combined together with the stochastic field theory bring us a way to understand and explain several complex phenomena in critical systems in clear and unified manner.  In fact, by the use of those methodologies, it has been found a variety of universal classes in the non-equilibrium systems, for instance, Kardar-Parisi-Zhang (KPZ) universality, directed percolation (DP) universality, parity conserving generalized voter (PCGV) universality, and pair contact process with diffusion (PCPD) universality, etc. 

In the context of ergodicity breaking, It is worth investing subject in which BRST symmetry plays a role in the stochastic processes and critical dynamics, and it is interesting to consider the RG flows towards the broken phase of non-equilibrium systems, where the basic properties in the macroscopic systems, dilation invariance or the ergodicity, would be broken.

In general, the correlation function at the critical point behaves as follows
\be
\left< \phi(x_1) \phi(x_2) \right> = \frac{1}{|x_1 - x_2|^{d-2 + \eta}} \;,
\ee
where $d$ stands for the dimension of the system, and $\eta$ is the so-called, critical exponent. There exists another interesting example in the case when $d=2$, then the large distance behavior of the correlation function at the critical point behaves like logarithmic divergent. The details are explained in the appendix.
\be
\left< \phi(x_1) \phi(x_2) \right> = \frac{1}{2 \pi} \log (|x_1 - x_2|) \;.
\ee
Therefore, the fluctuations become larger and larger, in other words, the correlations become stronger and stronger when the distance becomes larger and larger. That is, it becomes completely disordered and the mechanism behind this phenomena can be understood as a result of spontaneous breaking of global conformal symmetry ${\mathrm{SO}}(3,1) \simeq {\mathrm{SL}}(2,{\mathbb{C}})$, and which causes the appearance of massless Nambu-Goldstone (NG) mode, corresponds to the disordered phase for the fluctuating field $\phi$. In general, the conformal symmetry in $d$-dimensions is written by  ${\mathrm{SO}}(d+1,1)$, which is a combination of both dilatation (or, scale invariance) and rotation invariance. 

As the dilatation or scale invariance is related to the fundamental property of ergodicity which is discussed in the later suction, it is a relevant property which occurs in several places, for instance, scale invariance and universality appear near the phase transitions or the critical points. The property of scale invariance for the function $f: {\mathbb{R}}^{d} \to {\mathbb{C}}$ with the dilation $x \to \lambda x ~~(x \in {\mathbb{R}}^{d})$ is written by
\be
f (\lambda x ) = \lambda^{\Delta} f(x) \;,
\ee
for any dilations $\lambda \in {\mathbb{R}}$ and some choice of exponent $\Lambda$. Those class of functions $f$ is also said to be the homogeneous function with degree $\Delta$. In the case of stochastic processes, depending on the choice of scaling dimention $\Lambda$, probability distribution has a property
\be
{\mathbb{P}}(\lambda k) = \lambda^{\Delta} {\mathbb{P}}(k) \;,
\ee
where $k$ represents the frequency or the momentum, and it is said that $\Delta=0$ for white noise, $\Delta=-1$ for pink noise, and $\Delta=0$ for Brownian noise. 

It is well known fact that the gauge theory consists of the most fundamental element in quantum field theory. Even though the perturbative expansion in quantum field theory is so powerful in practice, it is necessary to treat the gauge theory beyond the perturbative expansion. At the root of the problem is so called, Gribov problem, or more generally the Gribov-Singer ambiguity. The main problem begins with the realizations that the local conditions do not have unique solutions beyond perturbation theory. There exist additional solutions, which are called as Gribov copies. However, it always exists the Gribov-Singer ambiguities as the equivalent classes of physical configurations are related by the gauge transformations. In other words, suppose if the system possesses some underlying there gauge symmetry, there exist some gauge redundancies, in which the distinct configurations in the theory are equivalent classes of field configurations. It is useful to consider the whole space of field configurations, curve which consist of gauge orbit do not intersect each other, and the hyper-surface spanned by the local gauge fixing constraint at the level of perturbation theory cut every curve orthogonally. Then, the Gribov copies arise because the curves cut the hyper-surface multiple times. Besides the first Gribov region, where the Faddeev-Popov determinant is positive semi-definite, the remainder of the curves of gauge orbit form a set of more Gribov region whose boundary could have zero or negative eigenvalues in the Faddeev-Popov determinant. 

In the stochastic field theory in the path integral formulation, the Gribov problem becomes more apparent. It is because the path integrals are defined in the entire field configurations, however, the path integrals need to be made over the physical configurations. In the stochastic approach, instead of considering the conditions for selecting a Gribov copy, random choice of a Gribov copy is taken for each residual gauge orbit. Assuming this approach can be done in such a way that the random selection of Gribov copy is made to be ergodic, unbiased, and well-defined, this is almost equivalent to the average over the residual gauge orbit. It still remains a Gribov problem in the path integrations.

The correct prescription to the Gribov problem has been founded by \citet{Kugo1} and \citet{Kugo2}, where the correct prescriptions for extracting the equivalent classes along the refinement of the BRST formalism has been formulated. In their formulation which remove the Gribov-Singer ambiguities or gauge redundancies, the physical state is defined as an element of cohomology of a BRST operator $Q_{B}$. That is, $\left|{\mathrm{Phys}} \right> \in {\mathrm{Ker}}(Q_{B})/{\mathrm{Im}}(Q_{B})$. One more comment is in order. Along the lines of Kugo-Ojima formalism, \citet{Ojima} has made investigations in a wide range of phenomena. They emphasized that emergence of physical but gauge dependent classical modes can be interpreted as the result of spontaneously symmetry breaking of BRST symmetry in each gauge sector. We give more detailed investigations on the relation with ergodicity later section in this study.

\section{Aspects of Algebra and Measure Theory}
\subsection{$C^{*}$-Algebra, KMS States, and Tomita-Takesaki theorem}
In this section, we summarize basic notions and properties of $C^{*}$-algebra, KMS States, and Tomita-Takesaki modular theory. 

\begin{definition}[$C^{*}$-algebra on Hilbert space $\cal{H}$]
A $C^{*}$-algebra $\cal{A}$ is the algebra of bounded linear operators defined on a separable infinite-dimensional Hilbert space $\cal{H}$, whose operator norm for $A \in \cal{A}$ is given by $\| A \| = \sup \left\{ \| A(x) \|  \mid x \in \cal{H} \right\}$. The $C^{*}$-algebra ${\cal{A}}$ is said to be unital if it includes an identity operator ${\bf{1}}_{{\cal{A}}} \in {\cal{A}}$.
\end{definition}

\begin{definition}[Representation of $C^{*}$-algebra]
A representation $\pi$ of a $C^{*}$-algebra on a Hilbert space $\cal{H}$ is a $\star$-homomorphism $\pi: \cal{A} \to \cal{B}(\cal{H})$. Two representations $\pi_{1}: \cal{A} \to \cal{B}(\cal{H_{1}})$ and $\pi_{2}: \cal{A} \to \cal{B}(\cal{H_{2}})$ is said to be equivalent if there is a unitary operator $U: {\cal{H}}_{1} \to {\cal{H}}_{2}$ such that $U \pi_{1}(A) U^{-1} = \pi_{2}(A)$ for all $A \in \cal{A}$. If a representation $\pi$ is an isomorphism onto its image, the representation is said to be faithful, and  a representation $\pi$ is said to be cyclic if there exists $\xi \in \cal{H}$ such that $\{\pi(A) \xi  \mid \forall A \in \cal{A} \}$ is dense.
\end{definition}

\begin{definition}[State on $C^{*}$-algebra]
A state $\varphi$ on a $C^{*}$-algebra $\cal{A}$ is a positive definite functional $\varphi: \cal{A} \to \mathbb{C}$ such that
\be
\| \varphi \| = \sup \{ \varphi(A) \mid A \in {\cal{A}}  \} = 1 \;,   
\ee
and $\varphi (A^{\star} A) \geq 0$ for all $A \in {\cal{A}}$. If $\varphi (A^{\star} A) > 0$ for all $A \in {\cal{A}}$, the state is said to be faithful. Let $S({\cal{A}})$ be a set of states on a $C^{*}$-algebra $\cal{A}$, a state $varphi \in S({\cal{A}})$ is said to be pure state if, for $\lambda \in [0,1 ]$ and $\varphi_{1}, \varphi_{2}  \in S({\cal{A}})$, we have
\be
\varphi = \lambda \, \varphi_{1} + (1-\lambda) \, \varphi_{2} ~\implies ~  \varphi = \varphi_{1} = \varphi_{2} \;.
\ee
\end{definition}

\begin{theorem}[GNS construction]
Let $\cal{A}$ be a $C^{*}$-algebra on a Hilbert space $\cal{H}$ and $\varphi$ be a state on it, then there is a cyclic representation $\pi: \cal{A} \to \cal{B}(\cal{H})$ with a basis $\xi \in \cal{H}$ such that 
\be
\varphi(A) = \left< \xi \, | \, \pi(A) \xi \right>
\ee
for all $A \in \cal{A}$. If $\pi^{\prime}: \cal{A} \to \cal{B}(\cal{H^{\prime}})$ is another cyclic representation with the basis $\xi^{\prime} \in \cal{H^{\prime}}$ such that $\varphi(A) = \left< \xi^{\prime} \, | \,   \pi^{\prime}(A) \xi^{\prime} \right>$, then $\pi$ and $\pi^{\prime}$ is equivalent. 
\end{theorem}

\begin{definition}[Gelfand-Naimark representation]
Let $\cal{A}$ be a $C^{*}$-algebra on a Hilbert space $\cal{H}$ and $\Phi$ be a set of pure states of $\cal{A}$.  For each $\varphi \in \Phi$, let $\pi_{\varphi}$ be the GNS representation of $\cal{A}$ on the Hilbert space ${\cal{H}}_{\varphi}$, the Gelfand-Naimark representation is defined to be the representation
\be
\bigoplus_{\varphi \in \Phi} \pi_{\varphi}: {\cal{A}} \to \bigoplus_{\varphi \in \Phi} {\cal{H}}_{\varphi}
\ee
\end{definition}

\begin{theorem}[Gelfand-Naimark theorem]
Let $\cal{A}$ be a $C^{*}$-algebra on a Hilbert space $\cal{H}$, then the Gelfand-Naimark representation is a faithful and cyclic representation of $\cal{A}$. If $\cal{A}$ is commutative $C^{*}$-algebra, then $\cal{A}$ is isometrically isomorphic to $C(X)$, the continuous functions on a locally compact Hausdorff space $X$, ${\it i.e.}$, ${\cal{A}} \simeq C(X)$. Hence, a state $\varphi$ on ${\cal{A}}$ provides a unique probability measure, Baire measure, on $X$ with
\be
\varphi(f) = \int_{X} f d \mu ~~\text{and}~~\mu(X) = \varphi(1) = 1 \;.
\ee
\end{theorem}

\begin{definition}[Strong and weak operator topologies]
Let ${\cal{B}}(\cal{H})$ be a set of bounded linear operators on a Hilbert space $\cal{H}$ and $\{T_{n}  \}  \subset {\cal{B}}(\cal{H})$ be a sequence. If $\| T(x) \to T_{n}(x) \| \to 0$ for all $x \in \cal{H}$, then $T_{n} \to T$ is said to be in the strong operator topology. If $\| \varphi(T(x)) \to \varphi(T_{n}(x)) \| \to 0$ for all $x \in \cal{H}$ and for all $\varphi \in {\cal{H}}^{\star}$, then $T_{n} \to T$ is said to be in the weak operator topology.
\end{definition}

\begin{definition}[von Neumann algebra ($W^{\star}$-algebra)]
A von Neumann algebra (or $W^{\star}$-algebra) ${\cal{M}}$ on a Hilbert space $\cal{H}$ is a $\star$-subalgebra of ${\cal{B}}(\cal{H})$ containing the identity operator, which is closed in the weak operator topology.
\end{definition}

\begin{definition}[Center of von Neumann algebra]
Let ${\cal{M}}$ be a von Neumann algebra. The commutant of a subset ${\cal{S}} \subseteq {\cal{M}}$ is defined by
\be
{\cal{S}}^{\prime} = \left\{ M \in {\cal{M}} \mid [ M,  S]  = 0 ~~\forall S \in {\cal{S}} \right\} \;. 
\ee
Furthermore, we define the double commutant ${\cal{S}}^{\prime \prime}$ of ${\cal{S}} \subseteq {\cal{M}}$ to be the commutant of the commutant, {\it i.e.}, ${\cal{S}}^{\prime \prime} = ({\cal{S}}^{\prime})^{\prime}$. The center of a subset ${\cal{S}} \subseteq {\cal{M}}$ is defined to be the set $\mathfrak{Z}({\cal{S}})$ of elements in ${\cal{S}}$ which commute with all elements from ${\cal{S}}$. That is,
\be
\mathfrak{Z}({\cal{S}}) = \left\{ S \in {\cal{S}} \mid [ S,  S^{\prime}]  = 0 ~~\forall S^{\prime} \in {\cal{S}} \right\} \;. 
\ee
The centre of a subset ${\cal{S}} \subseteq {\cal{M}}$ can be rewritten as $\mathfrak{Z}({\cal{S}}) = {\cal{S}} \cap  {\cal{S}}^{\prime}$.
\end{definition}

\begin{theorem}[von Neumann's double commutant theorem]
Let ${\cal{M}}$ be a subset of bounded linear operators on a Hilbert space ${\cal{H}}$ which contains the identity. Then the following conditions are equivalent. 
\begin{itemize}
\item $\begin{aligned}
{\cal{M}}~\text{is equal to the double commutant}~{\cal{M}}^{\prime \prime} \in {\cal{B}}({\cal{H}}), {\textit{i.e.},}~{\cal{M}} = {\cal{M}}^{\prime \prime}
\end{aligned}$
\item $\begin{aligned}
{\cal{M}}~\text{is a von Neumann algebra on}~{\cal{H}}~\text{with}~{\mathbf{1}}_{{\cal{B}}({\cal{H}})} = {\mathbf{1}}_{{\cal{M}}}
\end{aligned}$
\end{itemize}
If either of these conditions hold, then ${\cal{M}}$ is closed in the strong operator topology.
\end{theorem}

\begin{definition}[Normal state on von Neumann algebra]
Let ${\cal{M}}$ be a von Neumann algebra and $\varphi$ be a positive linear functional on ${\cal{M}}$. $\varphi$ is said to be normal if
\be
\varphi (\sup_{\alpha} A_{\alpha}) = \sup_{\alpha} \varphi (A_{\alpha})
\ee
for all increasing nets $\{ A_{\alpha} \}$ in ${\cal{M}}_{+}$ (${\cal{M}}_{+}$ is a set of positive elements in ${\cal{M}}$) with an upper bound.  
\end{definition}

\begin{definition}
Let $X$ be a topological space, $X$ is said to be separable of it contains a countable dense subset. If $X$ is a Hilbert space ${\cal{H}}$, ${\cal{H}}$ is said to be separable if and only if it has a countable orthonormal basis. Let ${\cal{H}}$, ${\cal{H}}$ be a separable Hilbert space, a bounded linear operator $A \in {\cal{B}}(\cal{H})$ is said to be trace-class if the following sum of positive elements becomes finite.
\be
\tr (\| A \|) = \sum_{j=1}^{n} \left< e_{j} \, | \, (A^{\star} A)^{1/2} \, e_{j} \right> \;.
\ee
\end{definition}

\begin{lemma}
Let $\varphi$ be a state on a $C^{*}$-algebra ${\cal{A}} = {\cal{B}}(\cal{H})$ on a Hilbert space $\cal{H}$. Then $\varphi$ is normal if and only if there exists a density matrix $\rho$, ${\it i.e.}$, a positive trace-class operator $\rho \in {\cal{C}}_{1}(\cal{H})$ with $\tr(\rho) = 1$, such that
\be
\varphi (A) = \tr_{\cal{H}} (\rho A) \;,~~ \forall A \in {\cal{B}}({\cal{H}}) \;.
\ee
In the case when we consider a finite dimensional Hilbert space ${\cal{H}}_{\Lambda}$, where $\Lambda$ is defined to be $\Lambda = {\mathbb{R}}^{d}$, the Gibbs state is normal
\be
\varphi_{\Lambda} (A) = \tr_{{\cal{H}}_{\Lambda}} (\rho_{\Lambda} A) \;,~~ \forall A \in {\cal{B}}({\cal{H}}_{\Lambda}) \;,
\ee
where the trace-class density operator is given by
\be
\rho_{\Lambda} = \frac{e^{-\beta H_{\Lambda} }}{ \tr_{{\cal{H}}_{\Lambda}} (e^{-\beta H_{\Lambda} }) } \;,~~ \forall A \in {\cal{B}}({\cal{H}}_{\Lambda}) \;.
\ee
\end{lemma}

\begin{theorem}[Maximal entropy theorem]
Let $M_{n}({\mathbb{C}})$ be a set of $n \times n$-matrices acting on ${\cal{H}}$. For a given state $\varphi \in {\cal{S}}(M_{n}({\mathbb{C}}))$ with the density operator $\rho$, von Neumann entropy is given by $S(\varphi) = - \tr [\rho \log \rho]$. Let $\varphi_{\beta}$ be a Gibbs state with the density operator $\rho_{\beta}$, the relative entropy is defined to be $S(\varphi \, | \, \varphi_{\beta}) = - \tr [ \rho \log \rho  - \rho \log \rho_{\Lambda} ]$. Then it satisfies the following inequality
\be
S(\varphi \, | \, \varphi_{\beta}) \leq S(\varphi ) \;,
\ee
and $S(\varphi \, | \, \varphi_{\beta}) =0$ if and only if $\varphi = \varphi_{\beta}$.
\end{theorem}

\begin{definition}[Modular operator]
Let ${\cal{M}}$ be a von Neumann algebra on a Hilbert space $\cal{H}$, $\Omega \in {\cal{H}}$ be a cyclic and separating vector of $\cal{H}$ with norm one. Here, the cyclic means that ${\cal{M}} \Omega$ is dense in $\cal{H}$, and the separating means that it satisfies the following property
\be
A \Omega = 0 ~\implies ~ A = 0
\ee
for all $A \in {\cal{M}}$. We can write a state vector $\varphi$ on ${\cal{M}}$ as $\varphi (A) = \left< \Omega \, | \, \pi(A) \Omega \right>$, based on the GNS construction. It is useful to define unbounded operators $S_{0}$ and $F_{0}$ on $\cal{H}$ such that
\beq
S_{0} (m \Omega) &=&  m^{\star} \Omega ~~\text{for all}~~m \in {\cal{M}} \\
F_{0} (m \Omega) &=&  m^{\star} \Omega ~~\text{for all}~~m \in {\cal{M}^{\prime}} 
\eeq
where ${\cal{M}^{\prime}}$ is the commutant of ${\cal{M}}$. We denote the closure of those operators as $S = \bar{S_{0}}$ and $F = \bar{F_{0}}$. Then they can be decomposed into polar decomposition,
\beq
S &=& J \Delta^{\frac{1}{2}} \ = \  \Delta^{-\frac{1}{2}} J  \\
F &=& J \Delta^{-\frac{1}{2}} \ = \ \Delta^{\frac{1}{2}} J 
\eeq
where $J = J^{-1} = J^{\star}$ is an anti-linear isometry of $\cal{H}$, called the modular conjugation, and $\Delta = F S$ is a positive, self-adjoint operator, called the modular operator. 
\end{definition}

\begin{theorem}[Tomita-Takesaki theorem]
Let ${\cal{M}}$ be a von Neumann algebra on a Hilbert space $\cal{H}$ with a cyclic and separating vector $\Omega \in {\cal{H}}$, $\Delta$ be the modular operator, and $J$ be the modular conjugation. It follows that
\beq
\Delta^{i t} {\cal{M}} \Delta^{- i t} &=& {\cal{M}} \\ 
J {\cal{M}} J &=& {\cal{M}^{\prime}} 
\eeq
for all $t \in {\mathbb{R}}$.
\end{theorem}

So far, the system described by neither the $C^{*}$-algebra nor the von Neumann algebra does not include the notion of time evolution.

\begin{definition}[$C^{*} (W^{*})$-dynamical system]
Let $\cal{A}$ be a unital $C^{*}$-algebra and $G$ be a locally compact group. Suppose that $\tau: G \to \text{Aut}(\cal{A})$ is a strongly continuous homomorphism, then the tuple $({\cal{A}}, G, \tau)$ is said to be a $C^{*}$-dynamical system. Similarly, let ${\cal{M}}$ be a von Neumann algebra and $G$ be a locally compact group. The tuple $({\cal{M}}, G, \tau)$ is said to be a $W^{*}$-dynamical system, if $\tau: G \to \text{Aut}(\cal{M})$ is a weakly continuous homomorphism.
\end{definition}

\begin{theorem}[Stone's theorem]
Let $U_{t}~(t \in {\mathbb{R}})$ be a strongly continuous one-parameter unitary group on a Hilbert space $\cal{H}$, then there exists a unique self-adjoint operator $A: D_{\cal{A}}  \subseteq {\cal{H}} \to {\cal{H}}$ such that
\be
U_{t} = e^{i t A}~~(\forall t \in {\mathbb{R}}) \,
\ee
where the domain of $A$ is given by
\be
D_{\cal{A}} = \left\{ \phi \in {\cal{H}} \mid - i \lim_{\epsilon \to 0} \frac{U_{\epsilon}(\phi) - \phi }{\epsilon} ~\text{exists}  \right\} \;.
\ee
Conversely, if $A: D_{\cal{A}}  \to {\cal{H}}$ is a self-adjoint operator on $D_{\cal{A}}  \subseteq {\cal{H}}$, then
\be
U_{t} = e^{i t A}~~(\forall t \in {\mathbb{R}}) 
\ee
is a strongly continuous one-parameter unitary group. 
\end{theorem}
Since the Hamiltonian $H$ is self-adjoint operator, the Stone's theorem supports that time evolution generated by $H$ is strongly continuous. The algebraic descriptions of a physical system are summarized as follows. A physical system is determined by a $C^{*}$-algebra of observables $\cal{A}$, the states of the physical system corresponds to the expectation value of that observable $A \in \cal{A}$ in the given state $\varphi$, ${\it i.e.}$, and the dynamics of a physical system is described by a $C^{*}$-dynamical system, in which a strongly continuous homomorphism $\tau_{t} = \tau(t)$ at every time $t \in {\mathbb{R}}$ determines the time evolution of each observables. 

\begin{theorem}[Modular group]
Let ${\cal{M}}$ be a von Neumann algebra and $\varphi$ be a faithful, normal state on ${\cal{M}}$. The Tomita-Takesaki theorem induces a one parameter group automorphism $\alpha: G \to \text{Aut}(\cal{M})$ such that 
\be
\alpha_{t} (M) = \pi_{\varphi}^{-1}  \left( \Delta^{i t} \pi_{\varphi}(M) \Delta^{- i t}  \right) \;,
\ee
for all $t \in G$ and $M \in {\cal{M}}$. Then the tuple $({\cal{M}}, G, \alpha)$ forms a $W^{*}$-dynamical system. It is called as the modular group automorphism. 
\end{theorem}

\begin{theorem}[Takesaki theorem]
Let $({\cal{A}}, {\mathbb{R}}, \tau)$ be a $C^{*}$-dynamical system, and $\varphi_{1}, \varphi_{2}$ be $(\tau, \beta_{1,2})$-KMS states for some $\beta_{1,2} \in {\mathbb{R}} \setminus \{ 0 \}$. Then each states $\varphi_{1}$ and $\varphi_{2}$ are disjoint, {\it i.e.}, $\varphi_{1}(A) \cap \varphi_{2}(A)  = \varnothing (\forall A \in {\cal{A}})$.
\end{theorem}

\begin{definition}[$(\tau, \beta)$-KMS state]
Let $({\cal{M}},{\mathbb{R}}, \tau)$ be a $W^{*}$-dynamical system, $\beta \in {\mathbb{R}}$, $\varphi$ be a state on $W^{*}$-algebra ${\cal{M}}$, we define a domain ${\cal{D}_{\beta}}$ such that
\[
{\cal{D}_{\beta}}
=
\begin{cases}
\{ z \in {\mathbb{C}}  \mid  0 < \Im z  < \beta  \} & \text{for}~~ \beta \geq 0  \\
\{ z \in {\mathbb{C}}  \mid  \beta < \Im z  < 0  \} & \text{for}~~ \beta < 0  
 \end{cases}
\]
and $\bar{{\cal{D}_{\beta}}}$ be the closure of ${\cal{D}_{\beta}}$ except for $\beta = 0$ where we set $\overline{{\cal{D}_{\beta}}} =  {\mathbb{R}}$. $\varphi$ is said to be a $(\tau, \beta)$-KMS state if it satisfies the following KMS conditions. That is, for every $A, B \in {\cal{M}}$ there exists a bounded continuous function $F_{A, B}: \overline{{\cal{D}_{\beta}}} \to {\mathbb{C}}$ analytic on ${\cal{D}_{\beta}}$ and such that for every $t \in {\mathbb{R}}$ it is true that
\beq
F_{A, B} (t) &=& \varphi(A \tau_{t}(B) ) \;, \\
F_{A, B} (t + i \beta ) &=& \varphi(\tau_{t}(B) A) \;.
\eeq
In the case when $\beta = -1$, $(\tau, -1)$-KMS state is called a $\tau$-KMS state.
\end{definition}

\begin{theorem}
Let $({\cal{A}}, {\mathbb{R}}, \tau)$ be a $C^{*}$-dynamical system and $\varphi$ be a $(\tau, \beta)$-KMS state for some $\beta \in {\mathbb{R}} \setminus \{ 0 \}$. Then for all $A \in {\cal{A}}$ and $t \in {\mathbb{R}}$, we have
\be
\pi ( \tau_{t}(A)) = \varphi (A) \;.
\ee
\end{theorem}

\begin{theorem}
Let $({\cal{A}(\cal{H})}, {\mathbb{C}}, \tau)$ be a $C^{*}$-dynamical system on a Hilbert space $\cal{H}$, and $\varphi$ be a $\beta$-Gibbs state which is defined by
\be
\varphi(A) = \frac{\tr (A e^{-\beta H} )}{\tr (e^{-\beta H})}
\ee
for all $A \in {\cal{A}}$, whose time evolution is given by
\be
\tau_{t}(A) = e^{i t H} A e^{- i t H} \;.
\ee
Then the state $\varphi$ on ${\cal{B}(\cal{H})}$ is a $\beta$-Gibbs state if and only if it is a $(\tau, \beta)$-KMS state. 
\end{theorem}

\begin{corollary}
Let $\varphi$ be a $(\tau, \beta)$-KMS state on a $C^{*}$-algebra. If $\varphi$ is a faithful state and ${\cal{A}}$ is commutative, then the one parameter group automorphism $\tau$ is trivial. Therefore, the $(\tau, \beta)$-KMS states can be regarded as the definition of the states in the equilibrium system. 
\end{corollary}

\begin{theorem}[Modular group and KMS state]
Let ${\cal{M}}$ be a von Neumann algebra and $\varphi$ be a faithful, normal state on ${\cal{M}}$. Then the tuple $({\cal{M}}, G, \alpha)$ is the unique $W^{*}$-dynamical system with respect to which $\varphi$ is an $\alpha$-KMS state where $\alpha$ is the modular group. 
\end{theorem}

\begin{corollary}
Let ${\cal{M}}$ be a von Neumann algebra and $\varphi$ be a faithful, normal state on ${\cal{M}}$. Then the tuple $({\cal{M}}, G, \tau)$ with $\tau_{t}(M) = \alpha_{-t/\beta}(M)$ and $\alpha$ the modular group of $({\cal{M}}, G, \varphi)$ is the unique $W^{*}$-dynamical system such that $\varphi$ is a $(\tau, \beta)$-KMS state.
\end{corollary}

\begin{definition}
Let $({\cal{M}},G, \tau)$ be a $W^{*}$-dynamical system, $\varphi$ be a faithful, normal state, and $\Omega$ be a cyclic vector for $\pi_{\varphi}({\cal{M}})$ in the Hilbert space ${\cal{H}}$. Then, with the use of the GNS representation, a one parameter group automorphism $\tau$ is said to be an inner automorphism if there is a unitary operator $U_{\varphi}(t)$ such that
\be
\pi_{\varphi}(\tau_{t}(M)) = U_{\varphi}^{\star}(t) \pi_{\varphi}(M) U_{\varphi}(t) ~~\text{with}~~U_{\varphi}(t) \Omega_{\varphi} = \Omega_{\varphi}
\ee
for all $M \in {\cal{M}}$ and $t \in G$.
\end{definition}

\begin{theorem}[Connes cocycle Radon-Nikodym theorem]
Let ${\cal{M}}$ be a von Neumann algebra and $\varphi_{1}$ and $\varphi_{2}$ be two faithful, normal states on ${\cal{M}}$. Then the modular group associated to $\varphi_{1}$ and $\varphi_{2}$ are related by an inner automorphism, {\it i.e.}, $\tau_{1}$ and $\tau_{2}$ are in the equivalent classes of automorphism denoted by ${\mathrm{Out}}({\cal{M}}) = {\mathrm{Aut}}({\cal{M}}) / {\mathrm{Inn}}({\cal{M}})$. 
\end{theorem}

Therefore, we can say that the time evolution of a physical system is determined by the modular automorphism group, associated to a state of a system, referred to as the thermal time $\beta = 1/T$. 

\begin{definition}[Ergodic state on von Neumann algebra]
Let $({\cal{M}}, G, \tau)$ be a $W^{*}$-dynamical system, in which a Lie-group $G$ acting on ${\cal{M}}$. We write $E_{{\cal{M}}}^{G}$ as a set of states, which is invariant under the action of $G$ on ${\cal{M}}$, ${\it i.e.}$, then the following equality holds with the use of the GNS construction, 
\be
\pi_{\varphi}(\tau_{g}(M)) = U^{\star}_{\varphi}(g) \pi_{\varphi}(M) U_{\varphi}(g)~~\text{with}~~U_{\varphi}(g) \Omega_{\varphi} = \Omega_{\varphi}
\ee
for all $M \in {\cal{M}}$ and $g \in G$. An ergodic state (or extremal point) is a state $\varphi \in E_{{\cal{M}}}^{G}$, which cannot be written as a proper convex combination of two distinct states $\varphi_{1}, \varphi_{2}  \in E_{{\cal{M}}}^{G}$. That is,
\be
\varphi \neq \lambda \, \varphi_{1} + (1-\lambda) \, \varphi_{2} ~~\text{unless}~~  \varphi = \varphi_{1} = \varphi_{2} 
\ee
for $\lambda \in (0,1)$.
\end{definition}

\begin{theorem}
Let $({\cal{M}}, G, \tau)$ be a $W^{*}$-dynamical system with an identity, and ${\cal{S}}_{\beta}({\cal{M}})$ be the set of all $(\tau, \beta)$-KMS states for $\beta > 0$. Then the following statements hold. 
\begin{itemize}
\item
The normal extension of $\varphi$ to $\pi_{\varphi}({\cal{M}})^{\prime \prime}$ is a $(\tau, \beta)$-KMS state.
\item
$\varphi \in {\cal{S}}_{\beta}({\cal{M}})$ is an ergodic state if and only if $\varphi$ is a factor state.
\item
The centre of ${\cal{M}}$, ${\mathfrak{Z}}({\cal{M}}) = {\mathbb{C}} {\bf{1}}$ consists of time invariant elements.
\item
If the GNS Hilbert space is separable, there exists a unique probability measure $\mu$ on ${\cal{S}}_{\beta}({\cal{M}})$, which is concentrated on the ergodic states such that
\be
\varphi(M) = \int_{E_{{\cal{M}}}^{G}} d \mu(\varphi^{\prime}) \varphi^{\prime}(M) \,,~~\text{for all}~~M \in {\cal{M}} \;.
\ee
\end{itemize}
Therefore, the central decomposition of a $(\tau, \beta)$-KMS state is identical to the extremal or the ergodic states decomposition. 
\end{theorem}

\begin{definition}[Spontaneous Symmetry Breaking]
Let $({\cal{M}}, G, \tau)$ be a $W^{*}$-dynamical system, in which a Lie-group $G$ is acting on ${\cal{M}}$, and a state $\varphi$ undergoes a Spontaneous Symmetry Breaking (SSB) of the group $G$ if the following conditions hold, that is, i) the state $\varphi$ is $G$-invariant, that is,
\begin{itemize}
\item
$\varphi(\tau_{g}(M)) = \varphi(M) \;,~~\forall g \in G,~\forall M \in {\cal{M}}$ 
\item
$\pi_{\varphi}(\tau_{g}(M)) = U^{\star}_{\varphi}(g) \pi_{\varphi}(M) U_{\varphi}(g)~~\text{with}~~U_{\varphi}(g) \Omega_{\varphi} = \Omega_{\varphi}$
\item
$U_{\varphi}(g) \pi_{\varphi}(M) \Omega_{\varphi} = \pi_{\varphi}(\tau_{g}(M)) \Omega_{\varphi}$
\end{itemize}
and ii) the state $\varphi$ has a non-trivial decomposition into ergodic states $\varphi^{\prime}$, {\it i.e.}, at least, two distinct states exist in the central decomposition of a $(\tau, \beta)$-KMS state with $\varphi^{\prime}(\tau_{g}(M)) \neq \varphi^{\prime}(M)$ for some $g \in G$ and for some $M \in {\cal{M}}$.
\end{definition}

\begin{corollary}
One of the simplest examples is the Gibbs state, which is invariant under the rotation $G = SO(d)$ ($d$ is spacial dimension). The ergodic state of this kind undergoes the spontaneous symmetry breaking from $G = SO(d)$ to the isometry subgroup $H = SO(d-1)$, whose degenerate vacuum corresponds to the harmonic space $G/H \simeq S^{d-1}$.
\end{corollary}

\subsection{Measure Equivalence, Cocycle, and Cohomology}
In this section, we discuss some aspects of group actions on the measure space, and its equivalent classes under the group actions.

\begin{definition}[Topological group]
A topological group $G$ is a topological space which has a group structure such that the group operation $(g, h) \in (G, G) \mapsto g \cdot h^{-1} \in G$ is continuous.
\end{definition}

\begin{definition}[Group action]
A group $G$-action on a measurable space $(X,\, {\mathcal{B}}, \mu)$, which is defined as a map $\alpha: G \times X \to X$, is said to be measure-preserving if for all $g \in G$, the map $x \mapsto g \cdot x ~ (x \in X, \; g \in G)$ is a measure-preserving isomorphism of $X$, {\it i.e.}, $\mu(g \cdot A) = \mu(A)\;,~\forall A \in {\mathcal{B}}$. We denote such a group action as $G \curvearrowright (X, \mu)$.
\end{definition}

\begin{definition}[Haar measure]
A Haar measure on a locally compact group $G$ is a Radom measure $\mu$ on $G$ which is invariant under left-translation, that is, for all $g \in G$ and for all Borel set $A \subseteq G$, $\mu(g A) = \mu(A)$. Here, the Radon measure $\mu$ is a Borel measure which is finite on compact and regular in the sense that for all Borel set $A$,
\be
\mu(A) = \sup \left\{ \mu(K) \mid K \subseteq A~\text{: compact}~ \right\} = \inf \left\{ \mu(U) \mid U \supseteq A~\text{: open set}~ \right\} \;.
\ee
\end{definition}

\begin{definition}[Measure equivalence]
Two infinite discrete countable groups $G$ and $\Lambda$ are said to be measure equivalent if there exists an infinite measure space $(X, \mu)$ with the measure-preserving actions $G \curvearrowright (X, \mu)$ and $\Lambda \curvearrowright (X, \mu)$ such that they admit finite measure fundamental domains $A, B \subset X$:
\be
X= \bigsqcup_{g \in G} g A =  \bigsqcup_{\lambda \in \Lambda} \lambda B \;.
\ee
The measure space $(X, \mu)$ is said to be a $(G, \Lambda)$ coupling.
\end{definition}

\begin{definition}[Quasi-invariant measure]
Let $(X,\, {\mathcal{B}}, \mu)$ be a measurable space, a measurable function $T: X \to X$ is said to be a measure-preserving transformation if it preserves the measure for all $A \in {\mathcal{B}}$. 
\be
\mu (T^{-1}(A)) = \mu(A) \;, ~~\forall A \in {\mathcal{B}} \;,
\ee
that is, it is written by a push-forward $T$ as $T_{\star}(\mu) = \mu$. Such a measure is called a quasi-invariant measure. Moreover, $\mu$ is said to be an ergodic measure if there are no $T$-invariant subsets up to measure 0. In this case, the Borel equivalence relation ${\mathcal{R}}$ is called the type-II equivalence relation.
\end{definition}

\begin{definition}[Ergodic action]
A measure-preserving group action $G \curvearrowright (X, \mu)$ is said to be ergodic if any $G$-invariant subset $A \subset X$ is null or co-null, {\it i.e.}, $\mu(A) = 0$ or $\mu(A \backslash E) = 0$. The action $G \curvearrowright (X, \mu)$ is said to be essentially free if for any $x \in X$ the stabilizer subgroup $G_{x} = \{ g \in G \mid g \cdot x = x  \}$ is trivial. 
\end{definition}

\begin{definition}[Countable equivalence relation]
Let $(X, \mu)$ be a standard Borel space and ${\mathcal{R}} \subset X \times X$ be a Borel subset which defines a countable Borel equivalence relation, where the equivalent classes are defined by ${\mathcal{R}}[x] = \left\{ y \in X \mid (x,y) \in {\mathcal{R}} \right\}$.
\end{definition}

\begin{definition}[Orbit equivalence relation]
Two group actions $G_{i} \curvearrowright (X_{i}, \mu_{i})~(i=1,2)$ are said to be an orbit equivalence relation if there is a measured space isomorphism $f: X_{1} \to X_{2}$ that sends orbits to orbits, that is,
\be
f(g_{1} \cdot x) = g_{2} \cdot f(x) \,,~\forall x \in X_{1} \;.
\ee
The orbit equivalence relation can be expressed by a countable Borel equivalence relation if $G \curvearrowright (X, \mu)$ is a 
\be
{\mathcal{R}}_{G \curvearrowright X} = \left\{ (x, \,g \cdot x)  \mid x \in X , g \in G  \right\} \;.
\ee
\end{definition}

\begin{theorem}[Feldman-Moore]
Any countable equivalence relation ${\mathcal{R}}$ on $(X, \mu)$ is the orbit equivalence relation ${\mathcal{R}}_{G \curvearrowright X}$ of a countable group $G$.
\end{theorem}

\begin{definition}[Full group]
Let $(X, \mu)$ be a standard Borel space. For a given countable equivalence relation ${\mathcal{R}}$, the full group $[ {\mathcal{R}} ]$ is defined by
\be
[ {\mathcal{R}} ] = \left\{T  \in {\mathrm{Aut}}(X, {\mathcal{B}}) \mid \forall x \in X\,, (x, T(x) )  \in {\mathcal{R}} \right\} \;.
\ee
Furthermore, let $A, B \subset X$ be two Borel subsets, a Borel isomorphism $T: A \to B$, where $A = {\mathrm{dom}}(T)$ and $B = {\mathrm{im}}(T)$, is called a partial isomorphism. the pseudo-Full group $[ [ {\mathcal{R}} ] ]$ is defined by
\be
[ [ {\mathcal{R}} ] ] = \left\{T  \in {\mathrm{Hom}}(A, B) \mid \forall x \in A \subset X\,, (x, T(x) )  \in {\mathcal{R}} \right\} \;.
\ee
\end{definition}

\begin{theorem}[Dye's reconstruction theorem]
Two ergodic countable equivalence relations ${\mathcal{R}}_{1}$ and ${\mathcal{R}}_{2}$ on $(X, \mu)$ are orbit equivalence if and only if their full groups are algebraically isomorphic. 
\end{theorem}

\begin{definition}[Restriction of equivalence relation]
Let ${\mathcal{R}}$ be a countable equivalence relation on $(X, \mu)$. For a given Borel subset $A \subset X$, restriction (or induction) of a equivalence relation ${\mathcal{R}}$ on $A$ denoted by ${\mathcal{R}}_{A}$ is defined by
\be
{\mathcal{R}}_{A} = {\mathcal{R}} \cap (A \times A) \;.
\ee
The restricted measure $\mu|_{A}$ is defined by $\mu|_{A}(E) = \mu(A \cap E)$. If $\mu$ is a probability measure, we denote by $\mu_{A}$, the normalized restriction $\mu_{A} = \mu(A)^{-1} \cdot \mu|_{A}$. 

For given subsets of equivalent classes ${\mathcal{R}}^{n}$ of ${\mathcal{R}}$, which is a Borel equivalence relation, ${\mathcal{R}}^{n}$ is said to be an increasing approximation if it holds ${\mathcal{R}} = \cup_{n} {\mathcal{R}}^{n}$. Moreover, a countable equivalence relation ${\mathcal{R}}$ is said to be hyperfinite if it admits an increasing approximation by finite subrelations. 
\end{definition}

\begin{definition}[Amendability]
Let ${\mathcal{R}}$ be a countable equivalence relation on $(X, \mu)$ with $\mu$ be an ${\mathcal{R}}$-quasi-invariant measure. Suppose $E$ be a separable Banach space, $C$ a measurable map $C: {\mathcal{R}} \to {\mathrm{Hom}}(E, E)$ is called 1-cocycle if it satisfies the following property: 
\be
{\cal{C}}(x, y) \, {\cal{C}}(y, z) = {\cal{C}}(x, z) \;.
\ee
The equivalence relation ${\mathcal{R}}$ is called amendable if it contains a measurable assignment $X \ni x \mapsto p(x) \in Q_{x} \subset E^{\star}$, so that
\be
{\cal{C}}(x, y)^{\star} p(x) = p(y) \;. 
\ee
\end{definition}

\begin{theorem}[Dye]
All the ergodic hyperfinite equivalence relations are mutually orbit equivalent.
\end{theorem}

\begin{theorem}[Conne-Feldman-Weiss]
All the amendable relations are hyperfinite. In fact, it can be generated by an action of ${\mathbb{Z}}$. We denote the amendable relation by ${\mathcal{R}}_{\mathrm{amen}}$.
\end{theorem}

\begin{definition}[Strongly Ergodic]
Let ${\mathcal{R}}$ be an ergodic equivalence relation on $(X, \mu)$. ${\cal{R}}$ is called strongly ergodic if every almost invariant sequence of Borel subset $A_{n} \subset X \,(n \in {\mathbb{N}})$ is trivial, {\it i.e.}, $\lim_{n \to \infty} \mu(A_{n}) \mu(A_{n} \backslash E) = 0$. 
\end{definition}

\begin{definition}[Fundamental group]
Let ${\mathcal{R}}$ be an equivalence relation on $(X, \mu)$. The fundamental group ${cal{F}}({\cal{R}})$ of the equivalence relation ${\cal{R}}$ is a subgroup of ${\mathbb{R}}_{+}^{\star}$ defined by
\be
{\cal{F}}({\cal{R}}) = \left\{ t \in {\mathbb{R}}_{+}^{\star} \mid {\cal{R}} \cong {\cal{R}}^{t}  \right\} \;.
\ee
Equivalently, the fundamental group ${cal{F}}({\cal{R}})$ consists of all ratios $\mu(A)/\mu(B)$ where $A, B \subset X$ are positive measure subsets of $X$ with ${\cal{R}}_{A} \cong {\cal{R}}_{B}$. Since the restriction of the amendable relation to any positive measure subset $A \subset X$ is amendable, it follows that
\be
{\cal{F}}({\cal{R}}_{\mathrm{amen}}) = {\mathbb{R}}_{+}^{\star} \;.
\ee
\end{definition}

\begin{definition}[Unitary representation of group]
Let $G$ be a topological group. A strongly continuous unitary representation of group $G$ on the Hilbert space $\cal{H}$ is a group homomorphism from $G$ to the unitary group of $\cal{H}$, {\it i.e.}, $\pi: G \to {\cal{U}}({\cal{H}})$ such that for all vector $\xi \in {\cal{H}}$, the map $g \in G \mapsto \pi(g) \xi \in {\cal{H}}$ is continuous. 
\end{definition}

\begin{definition}[Koopman representation]
Let us consider a measure-preserving group action $G \curvearrowright (X, \mu)$, a unitary representation of $G$ on $L^{2}(X, \mu)$ is said to be a Koopman representation (or regular representation) of $G$ if it holds the following property.
\be
\pi_{g} (f): x \in X  \mapsto f (g^{-1} x) \in L^{2}(X, \mu) \;, ~\forall g \in G\;, \forall f \in L^{2}(X, \mu) \;.
\ee
\end{definition}

\begin{corollary}
Let us consider a measure-preserving group action $G \curvearrowright (X, \mu)$, then the action is ergodic if and only if the only invariant functions $f \in L^{2}(X, \mu)$ under the Koopman representation are the constant functions.
\end{corollary}

\begin{lemma}
The amendable group action of ${\mathbb{Z}}$, equivalently, the rotation $R_{a}: x \mapsto x + a$ on the circle ${\mathbb{R}}/{\mathbb{Z}}$ is ergodic if and only if $a \in {\mathbb{R}}$ is irrational. Such a rotational transformation is closely related to the so-called, Fourier transformation $\cal{F}$, which relates $L^{2}({\mathbb{R}}/{\mathbb{Z}})$ with $\ell^{2}({\mathbb{Z}})$ as follows.
\be
{\cal{F}}: f \in L^{2}({\mathbb{R}}/{\mathbb{Z}}) \mapsto (c_{n}(f))_{n} \in \ell^{2}({\mathbb{Z}}) \;,
\ee
where $c_{n}(f) = \int_{0}^{1} dt f(t) \exp(- 2 \pi i n \, t)$. Then, the rotational transformation can be identified with the following transformation.
\be
T_{a}: (c_{n})_{n} \in \ell^{2}({\mathbb{Z}}) \mapsto \left(\exp( - 2 \pi i n \,a ) c_{n} \right)_{n}   \in \ell^{2}({\mathbb{Z}}) \;.
\ee
\end{lemma}

\begin{definition}[Amendable group]
Let $\pi: G \to {\cal{U}}({\cal{H}})$ be a unitary representation of a topological group $G$. Given a subset $K \subset G$ and $\epsilon > 0$, we say a vector $\xi \in {\cal{H}}$ is $(K, \epsilon)$-almost invariant if $\| \xi - \pi(g) \xi \| < \epsilon$ for all $g \in K$. A unitary representation $\pi$ of $G$, which has $(K, \epsilon)$-almost invariant vectors for all $K \subset G$ and $\epsilon > 0$, is said to weakly contain the trivial representation ${\bf{1}}_{G}$, denoted by ${\bf{1}}_{G}  \prec \pi$. A trivial representation ${\bf{1}}_{G}$ is strongly contained in $\pi$, denoted by ${\bf{1}}_{G}  < \pi$, if there exist non-zero $\pi(G)$ invariant vectors. 
\begin{itemize}
\item
A topological group $G$ is said to be amendable if the trivial representation of $G$ is weakly contained in the regular representation: $\pi: G \to {\cal{U}}(L^{2}(G))$, $\pi_{g} (f)(x) = f(g^{-1} x)$. 
\item
$G$ is said to have property (T) if for every unitary representation of $G$, ${\bf{1}}_{G}  \prec \pi$ implies ${\bf{1}}_{G}  < \pi$. This is equivalent to the existence of a subset $K \subset G$ and $\epsilon > 0$ so that any unitary representation $\pi$ of $G$ with $(K, \epsilon)$-almost invariant vectors, has non-trivial vectors. 
\end{itemize}
\end{definition}

\begin{definition}[Induced representation]
Let $(X, \mu)$ be a $(G, \Lambda)$ coupling and $\pi: \Lambda \to {\cal{U}}({\cal{H}})$ be a unitary representation of $\Lambda$ on the Hilbert space $\cal{H}$. Denote by ${\tilde{\cal{H}}}$ the Hilbert space consisting of equivalent classes (mod null-sets) of all measurable, $\Lambda$-equivalent maps $X \to {\cal{H}}$ with $\ell^{2}$-norm over the $\Lambda$-fundamental domain.
\be
{\tilde{\cal{H}}} = \left\{ f: X \to {\cal{H}} \mid f( \lambda \cdot x ) = \pi(\lambda) f(x) \;, \int_{X/\Lambda} \|f  \|^{2}  d \mu < \infty  ~~({\text{mod null-sets}})  \right\} \;.
\ee
The action of $G$ on such functions defines a unitary representation of $G$ as $\tilde{\pi}: G \to {\cal{U}}(\tilde{{\cal{H}}})$. This unitary representation is said to be induced from $\pi: \Lambda \to {\cal{U}}({\cal{H}})$ via $X$. 
\end{definition}

\begin{corollary}
Let $\pi: \Lambda \to {\cal{U}}({\cal{H}})$ be a unitary representation of $\Lambda$ on the Hilbert space ${\cal{H}}$, and $\tilde{\pi}: G \to {\cal{U}}(\tilde{{\cal{H}}})$ be the corresponding induced representation of $G$. Then, the following properties hold.
\begin{itemize}
\item
If $\pi$ is the regular unitary representation of $\Lambda$ on the Hilbert space ${\cal{H}} = \ell^{2}(\Lambda)$, then $\tilde{\pi}$ on ${\tilde{\cal{H}}}$ is identified with the unitary representation of $G$ over the space $L^{2}(X, \mu) \cong n \cdot \ell^{2}(G)$, where $n = {\mathrm{dim}}(L^{2}(X/\Lambda))$.
\item
If ${\bf{1}}_{\Lambda}  \prec \pi$ then ${\bf{1}}_{G}  \prec \tilde{\pi}$.
\end{itemize}
\end{corollary}

\begin{definition}[Cohomology group]
Let ${\cal{R}}$ be a countable equivalence relation on $(X, \mu)$. Denote ${\cal{R}}^{(n)}$ by
\be
{\cal{R}}^{(n)} = \left\{ (x_{0}, \dots , x_{n}) \in X^{n+1} \mid (x_{i}, x_{i+1}) \in  {\cal{R}}  \right\} 
\ee
equipped with the infinite Lebesgue measure ${\tilde{\mu}}^{(n)}$ defined by
\be
{\tilde{\mu}}^{(n)}(A) = \int_{X} {\#} \left\{ (x_{1}, \dots , x_{n}) \in X^{n} \mid   (x_{0}, \dots , x_{n}) \in  {\cal{R}}^{(n)}   \right\} d \mu(x_{0}) \;.
\ee
Take $({\cal{R}}^{(0)}, \tilde{\mu}^{(0)})$ to be $(X, \mu)$, then $({\cal{R}}^{(1)}, \tilde{\mu}^{(1)})$ is nothing but $({\cal{R}}, \tilde{\mu})$. Since $\mu$ is assumed to be ${\cal{R}}$-invariant, the above formula is invariant under the permutations of $(x_{0}, \dots , x_{n})$. 

Let us assume that $A$ is a compact Abelian group such as $A = {\mathbb{T}}$. The space ${\cal{C}}^{(n)}({\cal{R}}, A)$ of $n$-cochains consist of equivalence classes (mod null-sets) of measurable maps ${\cal{R}}^{(n)} \to A$, linked by the operators $d_{n}: {\cal{C}}^{(n)}({\cal{R}}, A) \to {\cal{C}}^{(n+1)}({\cal{R}}, A)$, defined by
\be
d_{n}(f)(x_{0}, \dots , x_{n+1}) = \prod_{i=0}^{n+1} f(x_{0}, \dots , \hat{x_{i}}, \dots , x_{0})^{(-1)^{i}} \;.
\ee
Note that $Z^{(n)}({\cal{R}}) = {\mathrm{Ker}}(d_{n})$ are the $n$-cocycles, and $B^{(n)}({\cal{R}}) = {\mathrm{Im}}(d_{n-1})$ are the $n$-coboundaries. The cohomology group is defined by $H^{(n)}({\cal{R}}) = Z^{(n)}({\cal{R}})/B^{(n)}({\cal{R}})$. In $n=1$ the $1$-cochains are the measurable maps ${\cal{C}}^{(1)}: ({\cal{R}}, \mu) \to A$ such that
\be
{\cal{C}}^{(1)}(x, y) \, {\cal{C}}^{(1)}(y, z) = {\cal{C}}^{(1)} (x, z) \;.
\ee
\end{definition}

\begin{corollary}
Let us consider a map $\varphi: G \to {\mathrm{Aut}}(G)$, from $G$ to the automorphism group of $G$ defined by $\tau_{g} = \tau(g)$, where $\tau_{g}$ is the automorphism of $G$ defined by
\be
\tau_{g}(h) = g \cdot h \cdot g^{-1} \;.
\ee
The function $\tau_{g}$ is said to be a group homomorphism, and its kernel is nothing but the center of $G$. and its image is said to be inner automorphism group of $G$ denoted by ${\mathrm{Inn}}(G)$. This means that
\be
G/Z(G) \simeq {\mathrm{Inn}}(G) \;.
\ee
The cokernel of this map is outer automorphism group of $G$ written by ${\mathrm{Out}}(G)$, and then this sequence form an exact sequence.
\be
1 \to Z(G) \to G \to {\mathrm{Aut}}(G) \to {\mathrm{Out}}(G) \to 1 \;.
\ee
\end{corollary}

\begin{definition}[Cocycle]
Let $G \curvearrowright (X, \mu)$ be a measure-preserving group action on $(X, \mu)$, and $H \subset G$ be a topological group. A Borel map: $C: G \times X \to H$ is said to be a cocycle if for every $g_{1}, g_{2} \in G$ and $x \in X$ it satisfies the following equality
\be
C(g_{2} \cdot g_{1}, x) = C(g_{2},  g_{1} \cdot x) \, C(g_{1}, x) \;.  
\ee
We denote the space of all cocycles by $Z^{1}(G \curvearrowright X, H)$. A cocycle which does not depend on the space variables $C(g, x) = C(g)$ naturally introduce homomorphisms: $C: G \to H$. 
\end{definition}

\begin{definition}[Cohomology space]
Let $\alpha: X \to H$ be a Borel map and then two cocycles $C_{1,2} \in Z^{1}(G \curvearrowright X, H)$ are said to be cohomologous (or equivalent), denoted by $C_{1} \sim C_{2}$, if there is a Borel map which holds the following equality
\be
C_{2} = \alpha(g \cdot x)^{-1}\, C_{1} \, \alpha(x) \;.
\ee
The space of equivalent classes (or quotient space) under the above identifications defines the cohomology space by
\be
H^{1}(G \curvearrowright X, H) = Z^{1}(G \curvearrowright X, H)/\sim \;.
\ee
A cocycle is said to be coboundary if it is cohomologous to the trivial cocycle, and we denote the set of coboundary by $B^{1}(G \curvearrowright X, H)$. 
\end{definition}

\begin{corollary}
If the measure-preserving group action acts by an abelian group $G$, then the cohomology space becomes the cohomology group: 
\be
H^{1}(G \curvearrowright X, H) = Z^{1}(G \curvearrowright X, H)/B^{1}(G \curvearrowright X, H) \;.
\ee
\end{corollary}

\begin{corollary}
If $\pi: (X, \mu) \to (Y, \nu)$ is an equivalent quotient map between $G$-actions ({\it i.e.}, $\pi_{\star}(\mu) = \nu$ and $\pi(g \cdot x) = g \cdot \pi(x)$ for $g \in G$), then for any target group $H$ any cocycle $C: G \times Y \to H$ lifts to another cocycle $\bar{C}: G \times X \to H$
\be
\bar{C}(g, x) = C(g, \pi(x)) \;, \forall x \in X \;.
\ee
Moreover, if $C^{\alpha} \sim C \in Z^{1}(G \curvearrowright Y, H)$, then the lifts $\bar{C}^{\alpha(\pi)} \sim \bar{C} \in Z^{1}(G \curvearrowright X, H)$, hence the map $X \xrightarrow{\pi} Y$ induces the map
\be
H^{1}(G \curvearrowright X, H)  \xleftarrow{~{\pi}^{\star}} H^{1}(G \curvearrowright Y, H) \;.
\ee 
\end{corollary}

\section{Stochastic Dynamics}
In this section, we first give a brief review of the field theoretical, path integral formulation of the stochastic dynamics. The action of a Langevin dynamics in a static equilibrium state possesses a time-scale translation symmetry. It leads to the Markov property of the Gibbs probability measure, which is independent of the field configurations. It is said that the scale invariance of probability measure is identical on the equivalence classes of the symmetry. If the ground state has a unique fields configuration for which the probability is maximized, those system is said to be ergodic. However, it is not the case if we consider the general situation with dynamics, the effective action strongly depends on the energy scale of interest and the scale invariance is spontaneously broken. Moreover, all systems are permanently interacting with their surroundings, which cause unavoidable fluctuations of extensive quantities, thus the systems become non-equilibrium as the time passes.

We give a brief description of statistical field theory in describing the dynamics of non-equilibrium systems. As usual in statistical field theory, we use the notation $x = (\tau, \overrightarrow{x}) \in \mathbb{R}^{d}$ with a Euclidean time $\tau = - i\, t $ with $\overrightarrow{x} \in \mathbb{R}^{d-1}$ and a parameter $\beta = 1/T$ stands for the inverse temperature. We use the effective field theory as a tool to describe and understand the stochastic dynamics in non-equilibrium systems. Specifically, the path integral formalism is used because it gives a deep understanding in which way to transform the measure under the symmetries. 

In a similar way to describe the stochastic differential equation in the normal spatial coordinates,
\be
dX = f(X) dt + g(X) dW(t) \;,
\ee
a functional form of stochastic differential equation takes the following form
\be
d\phi(x) = F[\phi] d\tau + K[\phi] dW(\tau) \;,
\ee
where $\phi(x)$ represents a stochastic random field, $F[\phi]$ is a differentiable functional with respect to $\phi$, and $K[\phi]$ is the Gaussian kernel depending on the field $\phi(x)$. By identifying the Brownian motion as $dW(\tau) = \eta(x) d \tau$ with an appropriate prescription. Then, the functional form of stochastic differential equation in the case of Langevin dynamics takes the following form
\be
\frac{\partial \phi(x)}{\partial \tau} = F[\phi] + K[\phi] \, \eta(x) \;,
\ee
where $\eta(x)$ is a Gaussian white noise. 

For a given observable ${\cal{O}}[\phi(x)]$, the expectation value of the observable ${\cal{O}}$ over the realizations of the Gaussian noise $\eta(x)$ is given by
\be
\left< {\cal{O}}[\phi] \right>_{\eta} = \int [{\cal{D}} \eta] {\mathbb{P}}[\eta] {\cal{O}}[\phi_{\eta}(x)] \;,
\ee
where $\phi_{\eta}(x)$ is the solution of Langevin equation for a given realization of the Gaussian noise $\eta(x)$, and $\mathrm{P}[\eta]$ is the probability distribution of the Gaussian noise $\eta(x)$. 

The probability distribution of the Gaussian noise $\eta(x)$ is given by the Gibbs measure
\be
{\mathbb{P}} [\eta] = \frac{1}{Z_0} \exp \left[ - S[\eta] \right] \;,
\ee
where $Z_0$ is a normalization constant. Here, the action for the Gaussian noise $\eta(x)$ is given by
\be
S[\eta] = \frac{1}{4} \int d^d x \, \bar{\eta}(x) \eta(x) \;.
\ee
It follows from the above definitions that the correlation function of the Gaussian noise $\eta(x)$ is determined to be
\be
\left<\bar{\eta}(x) \eta(x') \right> = 2 \delta^{(d)}(x-x') \;.
\ee
The expectation value of the observable ${\cal{O}}[\eta]$ can be written as follows
\beq
\left< {\cal{O}}[\phi] \right>_{\eta} &=& \int [{\cal{D}} \eta] {\mathbb{P}}[\eta] \int [{\cal{D}} \phi] \delta(G[\phi(x)]) {\cal{J}}[\phi] {\cal{O}}[\phi] \\
&=& \int [{\cal{D}} \eta] {\mathbb{P}}[\eta] \int [{\cal{D}} \phi] [{\cal{D}} \bar{\phi}] \exp \left[- \int d^d x \bar{\phi}(x) \left( \frac{\partial \phi(x)}{\partial \tau} - F[\phi] - K[\phi] \, \eta(x)  \right) \right]  {\cal{J}}[\phi] {\cal{O}}[\phi] \;,
\eeq
where $G[\phi(x)] = \frac{\partial \phi(x)}{\partial \tau} - F[\phi] - K[\phi] \, \eta(x)$, and ${\cal{J}}[\phi]$ is the Jacobian associated with the change of variables in the path integral. 
\beq
{\cal{J}}[\phi] &=& \Big| \det \left( \frac{\delta G[\phi]}{\delta \phi}   \right) \Big| \\
&=& \Big| \det \left( \partial \tau - \frac{\partial F[\phi]}{\partial \phi} - \frac{\partial K[\phi]}{\partial \phi} \, \eta(x) \right) \Big| \;.
\eeq
Hence, the result of integration over the Gaussian noise $\eta(x)$ becomes
\be
\left< {\cal{O}}[\phi] \right>_{\eta} = \int  [{\cal{D}} \bar{\phi}]  [{\cal{D}} \phi] \exp \left(- S[\bar{\phi}, \phi] \right) {\cal{J}}[\phi] {\cal{O}}[\phi]  \;,
\ee
where $S[\bar{\phi}, \phi]$ is the so called, Janssen-De Dominicis response functional that takes the following form.
\be
S[\bar{\phi}, \phi] = \int d^{d} x \bar{\phi} \left( \frac{\partial \phi}{\partial \tau} - F[\phi] - K[\phi]^2 \, \bar{\phi} \right) \;.
\ee
We still remain the evaluation of the Jacobian functional ${\cal{J}}[\phi]$, it depends on the choice of boundary conditions of the field $\phi(x)$ in general quantum field theory. However, in general, it requires careful prescriptions when we make change of field variables in the path integral formulation since the path integration does not care about the underlying symmetries in the action. One of the important constraints which need to be imposed in the process of path integrals is related to the BRST symmetry, in which the physical states are only invariant under the BRST transformation. 

In the case of Langevin dynamics, the constraint we have to impose can be written as follows
\be
\delta(G[\phi]) = \delta \left(\frac{\partial \phi}{\partial \tau} - F[\phi] - K[\phi] \, \eta \right) \;,
\ee
which is apparently divergent for the physical configurations. Therefore, it requires some careful prescriptions whenever the methods of field theoretical path integrations are used. 

Those constraints in the path integrals are very common issues in constrained systems in quantum field theory, the reason is because path integral is defined to be carried out all over the field configurations. The naive path integration would cause some redundancies since it contains not only the physical configurations but also unnecessary unphysical field configurations are included without any careful considerations. The physical configurations in field space respect the so called, BRST symmetry, by definition. The symmetries are not so manifest in the process of path integrations. It is not evident even if the unitarity of the theory, even if it is the most important principle in physics. 

There exists a procedure in which the symmetries are manifestly covariant by introducing the anti-commuting Grassmann variables or the Faddeev-Popov ghosts in the path integrals. The Faddeev-Popov ghosts do not have any corresponding physical states since it violate the spin-statistics relation. The Faddeev-Popov ghosts are the complex scalar fields which are anti-commuting, Grassmann variables
\be
\{ \bar{c}(x), c(x) \} = \delta(x-x') \;.
\ee
Since the Faddeev-Popov ghosts are introduced as the auxiliary fields, they do not co respond to any physical states. Hence, their boundary conditions in the path integrals need to be chosen as follows
\be
c(x) = \bar{c}(x) = 0 ~~({\text{for}}~x \to \pm \infty) \;.
\ee
The so-called, BRST operator $Q$ is defined for any fields $\phi $ to transform as
\be
\delta \phi = \epsilon Q \phi \;,
\ee
and it satisfies a nilpotent property
\be
Q^2 = 0 \;.
\ee
The BRST symmetry is the crucial property in field theories as the physical states in the field configurations is BRST invariant. Therefore, it can be used to define the relevant field configurations in the entire field configurations. Especially, it is so important to investigate on the relationship between the BRST symmetry and the ergodic symmetry, which is related to one of the crucial subjects in this study. 

Remember that what we need to consider is an appropriate prescription in evaluating the Jacobian ${\cal{J}}[\phi] = \Big| \det \left( \frac{\delta G[\phi]}{\delta \phi} \right) \Big|$, which appears in the path integral measure under the transformation of the fields. In general, it is inevitable for the appearance of additional Jacobian factor in path integrals. Therefore, it is necessary to take some appropriate prescriptions since the determinant inside the Jacobian becomes divergent without any considerations. One of the prescriptions in evaluating the Jacobian factor is to introduce the Faddeev-Popov ghosts. 

Suppose if $M$ is a complex valued commutative matrix, the integrals of the Faddeev-Popov ghosts satisfy the following equality
\be
\det (M) = \int [{\cal{D}} \bar{c}] [{\cal{D}} c]  \exp \left( \int d^d x  \bar{c}(x) \det (M) c(x) \right)  \;.
\ee
Then, we can write it down the expectation value of the observable ${\cal{O}}[\phi]$ as follows
\be
\left< {\cal{O}}[\phi] \right>_{\eta} = \int [{\cal{D}} \bar{\phi}] [{\cal{D}} \phi] [{\cal{D}} \bar{c}] [{\cal{D}} c] 
\exp \left(- S[\bar{\phi}, \phi, \bar{c}, c] \right) {\cal{O}}[\phi]  \;,
\ee
where the action $S=S[\bar{\phi}, \phi, \bar{c}, c] ]$ is given by
\be
S[\bar{\phi}, \phi, \bar{c}, c] ] = \int d^{d} x \left[ \bar{\phi} \left( \frac{\partial \phi}{\partial \tau} - F[\phi] - K[\phi]^2 \, \bar{\phi} \right)
+  c(x)  \left( \frac{\partial}{\partial \tau} - \frac{\partial F[\phi]}{\partial \phi} \right) \bar{c}(x) \right] \;.
\ee
The partition functional takes the following form in the same way as in the equilibrium systems
\be
Z[\bar{j}, j] = \int [{\cal{D}} \bar{\phi}] [{\cal{D}} \phi] [{\cal{D}} \bar{c}] [{\cal{D}} c]  
\exp \left[- S[\bar{\phi}, \phi, \bar{c}, c] + \int d^d x \left(\bar{j}(x) \bar{\phi}(x) + j(x) \phi(x) \right)  \right] \;.
\ee
It is well known fact that stochastic field theory which describes in general non-equilibrium system possesses an underlying symmetry, supersymmetry. In fact, when the noise kernel $K[\phi]$ is given by some Hamiltonian
\be
K[\phi(x)] = \frac{\delta {\cal{H}}}{\delta \phi(x)} \;,
\ee
then the stochastic dynamics converge to the equilibrium systems. 

Fortunately, in studying the stochastic processes in the scheme of the Ito's prescription, it has been proved that the Jacobian in the stochastic path integrals is independent of choice of fields in the path integrals, and hence we can choose to be ${\cal{J}}[\phi] = 1$. Now, we consider the effective action in the Langevin dynamics. The partition functional in the system can be written by
\be
Z[J] = \int [{\cal{D}} \Phi] \exp \left[ - S[\Phi] +  \int d^d x \, J^{\text{T}}(x) \Phi(x)]  \right] \;,
\ee
where a matrix notation is introduced for simplicity as
\be
\Phi(x) =
\begin{pmatrix}
\bar{\phi}(x) \\
\phi(x) 
\end{pmatrix} 
\;,
~~
J(x) =
\begin{pmatrix}
\bar{j}(x) \\
j(x)
\end{pmatrix}
\;.
\ee
The effective action can be calculated in the same way to the one in the statistical field theory in equilibrium systems
\be
\Gamma[\varphi] = \int d^d x \, J^{\text{T}}(x) \varphi(x) - W[J] \;,
\ee
where $W[J] = \log Z[J]$ and $\varphi(x) = \left< \Phi(x) \right>$ is the background field.

\section{Spontaneous Ergodicity Breaking}
Recently, ergodicity breaking has received a number of interests in a variety of fields related to the stochastic processes and the others. The concept of ergodicity is basically a measure property under the symmetries. There are several studies on the ergodicity breaking in a variety of fields. One of the arguments is that it causes less influences in most practical circumstances even if the ergodicity breaking is theoretically true. However, it is worth investigations to investigate on which conditions or underlying symmetries controls the ergodicity, in general. Furthermore, there could be some implications in studying the anomalous behaviors caused by the ergodicity breaking. 

Here we summarize the original statement of Birkhoff's ergodicity theorem in probability theory. It states that in a class of ergodic systems the time average of an observable converges to the average over the entire phase space in the dynamical system. Let us briefly summarize the Birkhoff's ergodicity theorem. It is supposed that $({\bf{X}},\, \bf{\Omega},\, \mu)$ is a probability space and ${\mathcal{T}}: X \to X~(X \in {\bf{X}})$ is measure preserving transformation. There are some examples of such a measure preserving transformation, for instance, a flow map in random dynamical systems, Hamiltonian flow in dynamical systems, and the others. For the later use, it is necessary to define successive transformations of a single measure preserving transformation. It is defined to be $T_{s}: X \to X~(X \in {\bf{X}})$ for $s \in {\mathbb{Z}}$, which obeys the following rules
\begin{itemize}
\item $\begin{aligned}
T_{0} (X) &=& X 
\end{aligned}$
\item $\begin{aligned}
T_{s} (T_{t} (X)) &=& T_{s+t} (X) 
\end{aligned}$
\item $\begin{aligned}
T_{t}^{-1} (X) &=& T_{-t} (X) 
\end{aligned}$
\end{itemize}
Let $f \in L^{1}({\bf{X}},\, \bf{\Omega},\, \mu)$, then the Birkhoff's ergodicity theorem (or, pointwise ergodic theorem) states that
\be
\lim_{n \to \infty} \frac{1}{n} \sum_{i=0}^{n-1} f(T_i(X)) = \int_{\Omega} f d \mu \;,
\ee
where $X \in {\bf{X}}$ and $\Omega \in {\bf{\Omega}}$. 

The Kac's lemma tells us the expected value of returning time in the ergodic process. Let us suppose ${\mathcal{T}}$ be an ergodic measure preserving transformation of the probability space $({\bf{X}},\, \bf{\Omega},\, \mu)$ and let $\bf{A} \subset \bf{X}$ with $\mu({\bf{A}}) > 0$, then, based on the Poincare Rucurrence theorem, we can define the retuning time as
\be
\tau_{{\bf{A}}}(X) =  \inf \{ n \geq 1 \mid T_{n}(X) \in {\bf{A}} \} \;,
\ee
for $X \in {\bf{A}}$. If ${\mathcal{T}}$ is invertible, then the Kac's lemma says that
\be
\int \tau_{{\bf{A}}}(X) d \mu = 1 \;.
\ee
In particular, 
\be
{\mathbb{E}} \left[\, \tau_{{\bf{A}}}  \mid {\bf{A}}\, \right] = \frac{1}{\mu({\bf{A}})} \;. 
\ee
It says that when the ergodic system have a trajectory starting from $A \in {\bf{A}}$ and returning back to $A \in {\bf{A}}$, whose expected time to return is given by $1/\mu({\bf{A}})$. Based on this obserbation for the retun period, as the phase space in concern becomes larger and larger, it takes shorrter and shorter period of time to converge the time average to the phase space average. On the contrary, as if the phase space in concern is so small compared to the size of the entire system, it takes so long period of time in order to converge the time average to the phase space average.

\newpage
\appendix
\section*{Appendix}
\section{From Langevin to Fokker-Planck}
In this Appendix, we give a summary of our notations by giving a brief review on the several models which are necessary in our study. Let us begin with the Langevin dynamics described as follows.
\be
\frac{\partial \phi(x)}{\partial \tau} = F[\phi] + K[\phi] \, \eta(x) \;.
\ee
Here, we suppose that $\varphi(x)$ being a solution of the above Langevin equation for which some particular trajectory of $\phi(x)$ kicked by the Gaussian noise $\eta(x)$. Then, the normalized probability distribution is given by 
\be
\rho [ \phi(x) ] = \delta( \varphi(x) - \phi(x)) \;.
\ee
By taking a time derivative, it becomes
\be
\frac{\partial \rho}{\partial \tau} = \frac{\partial }{\partial \phi} \left( \frac{\partial \phi}{\partial \tau}\Big|_{\phi = \varphi} \, \rho [\phi(x) ]   \right) \;,
\ee
and the solution is simply given by
\be
\rho [\phi(x) ]  = \exp \left( - \int d^d x \frac{\partial}{\partial \varphi} \frac{\partial \phi}{\partial \tau}\Big|_{\phi = \varphi}  \right) \rho[\phi(0)] \;. 
\ee
Now, let us define the probability distribution ${\mathrm{P}}[\phi(x) ] $ which is averaged over all trajectories of the Gaussian noise $\eta$. 
\be
{\mathbb{P}}[\phi(x) ]  = \left< \rho [\phi(x) ]  \right>_{\eta} \;,
\ee
which leads to the Fokker-Planck equation as below.
\be
\frac{\partial {\mathbb{P}}[\phi(x) ] }{\partial \tau} = \frac{\partial }{\partial \phi} \left( - F[\phi] + \frac{\partial K[\phi]}{\partial \phi}  \right) {\mathbb{P}}[\phi(x) ]  \;.
\ee

\section{Effective Field Theory}
The effective field theory is a class of effective theories describing the physics at a scale of interest $\mu$ with $\mu \ll \Lambda$, where $\Lambda$ is a UV cut-off of the effective field theory. When we could know the theory above the UV cut-off of the low-energy effective field theory, the low-energy effective field theory have to be constructed by integrating out the high-energy degrees of freedom. Such a scale transformation would change the values of coupling constants, namely the interaction strengths in the effective field theory. Those changes of coupling constants are described by the renormalization group (RG) equation.

The finite temperature field theory has been developed to calculate the expectation values of physical observables in quantum field theory at finite temperature. The generaling functional is given by the Euclidean path integral of the form:
\be
Z[J] = \int [{\cal{D}} \phi] \exp\left(- S[\phi] + \int d^dx J \cdot \phi  \right) \Big|_{\phi(\beta) = \phi(0)} \;,
\ee
where $S$ is the Euclidean action, $x$ is the Euclidean coordinates and $J$ is an auxiliary field which plays a role as a generating functional to calculate the $n$-point correlation functions (Green functions).
\be
\left<\phi_1, \cdots, \phi_n \right> = \frac{1}{Z[0]} \frac{\delta Z[J]}{\delta J_1 \cdots \delta J_n}\Big|_{J=0} \;.
\ee
Let us consider the $\phi^{4}$ scalar field theory, whose action is giveb by
\be
S[\phi] = \int d^dx \left[ \frac{1}{2} \left(\nabla \phi \right)^2 + \frac{1}{2} m^2 \phi^2
+ \frac{1}{4!} g \phi^4 \right] \;.
\ee
The connected generaling functional $W[J]$ is defined to be
\be
W[J] = \log Z[J] \;.
\ee
This is, in fact, the generating functional for the connected correlation functions
\be
\left<\phi_1, \cdots, \phi_n \right>^{(c)} = \frac{\delta W[J]}{\delta J_1, \cdots, \delta J_n}\Big|_{J=0} \;.
\ee
There is an alternative description for $W[J]$ as a series expansions.
\be
W[J] = \sum_{n \geq 0} \frac{1}{n!} \int d^dx_1 \cdots d^dx_n  \left<\phi_1, \cdots, \phi_n \right>^{(c)}
J_1 \cdots J_n \;.
\ee
In the case of Brownian free field, the generating function is written by
\be
Z[J] = \exp \left[ \frac{1}{2} \int d^dx d^dy J(x) G(x, y) J(y) \right] \;,
\ee
with $G(x, y)$ the Green function.

The effective action (or vertex operator) $\Gamma[\varphi]$ is also important functional in the effective field theory, which is defined by the Legendre transform of $W[J]$.
\be
\Gamma[\varphi ] = \int d^d x J(x) \varphi(x) - W[J] \;,~~ \varphi(x) = \frac{\delta W[J]}{\delta J(x)} \;.
\ee
The auxiliary field $\varphi(x)$ is called the background field. We can expand the effective action in power of $\varphi(x)$.
\be
\Gamma[\varphi] = \sum_{n} \frac{1}{n!} \int \prod_{i=1}^{n} \frac{d^d x_i}{(2 \pi)^d}
\hat{\varphi}(x_i) \Gamma^{(n)}(x_1, \cdots, x_n) \;,
\ee
where the function $\Gamma^{(n)}(x_1, \cdots, x_n)$ are called the $n$-point vertex functions. The effective action contains all the interactions generated through the original action in the effective field theory.

The effective action can be expressed in terms of the background field $\varphi(x)$ on the Fourier space as
\be
\varphi (x) = \int \frac{d^d k}{(2 \pi)^d} e^{i k x} \hat{\varphi}(k) \;,
\ee
and then, the effective action is written by
\be
\Gamma[\hat{\varphi}] = \sum_{n} \frac{1}{n!} \int \prod_{i=1}^{n} \frac{d^d k_i}{(2 \pi)^d}
\hat{\varphi}(-k_i) \Gamma^{(n)}(k_1, \cdots, k_n) \;.
\ee
Now we give a simple example to show the way to calculate the effective action.
\be
Z[J] = \int [{\cal{D}} \phi] e^{- S[\phi; J]} \;,~~S[\phi; J] = S[\phi] - (J, \phi) \;,
\ee
where we have introduced a notation, $(J, \phi) = \int d^d x J(x) \phi(x)$. It follows from the fact that the background field is a vacuum expectation value of the field.
\be
J(x) = \frac{\delta S[\varphi]}{\delta \varphi(x)} \;.
\ee
We can write down the generating functional for the connected correlation functions as
\be
W[J] = - S[\varphi] + (J, \varphi) - \frac{1}{2} \text{tr} \log \left[{\cal{H}} S[\varphi] \right] \;,
\ee
where ${\cal{H}}$ is the Hessian matrix, then the effective action is calculated by the Legendre transformation: $\Gamma[\varphi] = (J, \varphi) - W[J]$.
That is,
\be
\Gamma[\varphi]  = S[\varphi] +  \frac{1}{2} \text{tr} \log \left[{\cal{H}} S[\varphi] \right] \;.
\ee
Since the action in $\phi^4$ theory is given by
\be
S[\varphi] = \int d^d x \left[ \frac{1}{2} \left(\nabla \phi \right)^2 + \frac{1}{2} m^2 \phi^2+ \frac{1}{4!} g \phi^4   \right] \;,
\ee
the effective action in $\phi^4$ theory is calculated to be
\be
\Gamma[\varphi]  =  \frac{1}{2} \text{tr} \log \left[ - \nabla^2 + m^2 \right]
+ \frac{1}{2} \text{tr} \log \left[ 1 + \frac{g}{2} \left( - \nabla^2 + m^2 \right)^{-1} \varphi^2 \right]  \;.
\ee

Then, the perturbative expansion of the effective action as power series of $g$ can be written by
\be
\Gamma[\varphi]  =  \sum_{k=1}^{\infty} \frac{g^k}{2^k} \frac{(-1)^{(k+1)}}{k} \Gamma^{(k)}[\varphi] \;,
\ee
where the $k$-th term of the effective action in $\phi^4$ theory is given by
\be
\Gamma^{(k)}[\varphi] = \text{tr} \log \left[ \left( - \nabla^2 + m^2 \right) \varphi^2 \right]^k \;.
\ee
It can be expressed as a product of two point functions
\be
\Gamma^{(k)}[\varphi] = \int \prod_{i=1}^{k-1} d^d x_i \varphi^2(x_i) G(x_i - x_{i+1}) \varphi^2(x_k) G(x_k - x_1) \;.
\ee

For the constant background field, the effective action in $\phi^4$ theory reads
\be
\Gamma[\varphi] = S[\varphi] +  \frac{1}{2} \int \frac{d^d k}{(2 \pi)^d} \log \left[ k^2 + {\cal{H}} S[\varphi] \right] \;.
\ee
By taking the limit of $d \to 4$, the effective action in $\phi^4$ theory at the scale of interest $\mu \ll \Lambda$ is given by
\be
\Gamma [\varphi] = \frac{1}{2} m^2 \varphi^2 + \frac{1}{4!} g \varphi^4
+ \frac{1}{(8 \pi)^2} \left[ {\cal{H}} S[\varphi] \right]^2 \log \left[ \frac{{\cal{H}} S[\varphi]}{\mu^2} \right] \;,
\ee
When we rewrite the coupling constants in the classical action with subscript $0$, the running coupling constants read to be
\beq
m^2(\mu) &=&  m_0^2 +  \frac{g_0}{2 (4 \pi)^2} \left[ \Lambda^2 + m_0^2 \log \left( \frac{\mu^2}{\Lambda^2} \right)  \right] \;,  \\
g(\mu) &=& g_0 +  \frac{3 g_0^2 }{2 (4 \pi)^2} \log \left( \frac{\mu^2}{\Lambda^2} \right) \;.
\eeq
These scaling behavior of the effective action is the key concept of the effective field theory. In fact, the low energy effective action is constructed by integrating out the high-energy degrees of freedom in the original theory. It is related to the Renormalization Group (RG) transformation.

The RG flows of the coupling constants are determined by the beta-functions, which is defined by
\be
\beta(g) = \Lambda \frac{\delta g}{\delta \Lambda} \;.
\ee
In a simple case with only one coupling constant, solution to the beta function is simply given by
\be
g(\Lambda) = \exp \left[ \int \frac{dg'}{\beta(g')} \right] \;.
\ee
For example, the one-loop beta functions in $\phi^4$ theory are given by
\beq
\Lambda  \frac{\delta g}{\delta \Lambda} &=& \epsilon g - \frac{3}{(4 \pi)^2} g^2 \;, \\
\Lambda  \frac{\delta m^2}{\delta \Lambda} &=& 2 m^2 + \frac{g}{(4 \pi)^2} \left( 1 - m^2   \right) \;,
\eeq
where we have defined the theory in $d = 4 - \epsilon$ dimensions to make it finite. In this case, there exists a non-trivial Wilson-Fisher fixed point at
\be
g_* = \frac{16 \pi^2}{3} \epsilon \;, ~~ m_*^2 = - \frac{1}{6} \epsilon \;.
\ee
Near the Wilson-Fisher fixed point, there are two phases, one corresponds to the broken phase toward $m_*^2 \to - \infty$, and the other is the symmetric phase towards $m_*^2 \to + \infty$.

Here we summarize the notion of Green function (or propagator). The Green function is the kernel of the Laplace operator $\left(- \Delta + m^2 \right)$:
\be
\left(- \Delta + m^2 \right) G(x, x') = \delta(x, x') \;.
\ee
After the Fourier transition, the solution is given by
\be
G(x, x') = \int \frac{d^dk}{(2 \pi)^d} e^{i k \dot (x-x')} \frac{1}{\left(k^2 + m^2 \right)^2} \;.
\ee
When going back to the spacial dimensions with the use of radial coordinates, we have
\be
G(r) = \frac{S_{d-1}}{(2\pi)^{d-1}} \int \frac{dk}{2\pi} \frac{k^{d-1}}{k^2 + m^2}
\int_{0}^{\pi} d\theta (\sin \theta)^{d^2} e^{i k r \cos \theta} \;,
\ee
with $S_d = 2 \pi^{d/2} / \Gamma(d/2)$ the volume of the unit sphere in $\mathbb{R}^d$. Then, we have
\be
G(r) = \frac{1}{(2\pi)^{d/2}} \left(\frac{m}{r} \right)^{d-2} K_{\frac{d-2}{2}}(m r) \;.
\ee
The explicit form of this solution can be shown in $d=3$ and $d=2$. In the case of $d=3$, it can be expressed as
\be
G(r)_{d=3} = \frac{1}{4 \pi} \frac{1}{r} e^{- m r}  \;.
\ee

Now we consider two limiting situations with short and large scale behaviors. At short distances, we have the following expressions.
\beq
G(r)_{d=3} &=& \frac{1}{4 \pi} \frac{1}{r} + \cdots \;, \\
G(r)_{d=2} &=& \frac{1}{2 \pi} \log ( \frac{2}{m r}) \;.
\eeq
On the other hand, at large distances, the Green function has an exponential decrease factor if the mass term is non-vanishing.
\be
G(r)_{d} \sim \frac{1}{r^{d-2}} e^{- m r} \;.
\ee
This shows the correlation function is given by $\xi = 1/m$. If we take the limit $m \to 0$,
\be
G(r)_{d} = \frac{1}{(2 \pi)^{d}} \int dk d\Omega k^{d-3} e^{i k r \cos \theta} \sim  \frac{1}{r^{d-2}}  \;.
\ee
There is a logarithmic divergence in $d=2$, we thus have to introduce the momentum cut-off $\Lambda$.
\be
G(r)_{d=2} = \frac{1}{2 \pi} \log (\Lambda r) \;.
\ee

\section{Conformal Field Theory}
It is trivial in a sense that the Brownian theory without any quantum corrections has conformal invariance, and the two point function of a primary field ${\mathcal{O}}$ in the conformal field theory behaves like
\be
\left<{\mathcal{O}}(x_1) {\mathcal{O}}(x_2) \right> = \frac{1}{| x_1 - x_2 |^{\Delta}} \;,
\ee
where $\Delta$ is the conformal dimension of the primary field. The scaling dimension of the scalar field is given by $\Delta = d-2+\eta$ where $\eta$ is the anomalous dimension or critical exponent. $\Delta=1$ corresponds to the Wilson-Fisher fixed point in $d=4$ dimensions. Nonetheless, the scale invariance is broken, if we turn on the quantum corrections to the theory because the coupling constants vary with the distance scale according to the renormalization group equations. Hence the classical scale invariance (or conformal invariance) is said to be anomalous.

We continue to investigate on the massless scalar field theory in $d=2$. In fact, it gives the simplest example of $d=2$ CFT. It is useful to define the complex coordinates:
\be
z = x + i y \;, ~ \bar{z} = x - iy \;,
\ee
then the Euclidian metric is given by $ds^2 = g_{\mu \nu} dx^{\mu} dx^{\nu} = dx^2 + dy^2 = dz d\bar{z}$.

The action for the massless scalar field is given by
\be
S[\phi] = \frac{g}{8 \pi} \int dz d\bar{z} (\partial_z \phi)(\partial_{\bar{z}} \phi) \;.
\ee
The Green function of the $d=2$ Laplacian is given by
\be
G(z, w) = - \frac{1}{2 \pi} \log | z - w | \;,
\ee
which satisfies $(-\Delta_z) G(z, w) = \delta^{(2)} (z - w)$. The Green function is only determined up to a constant because the Laplacian possesses a constant zero mode. Hence, we have to deal with such a constant zero mode. One way to solve this problem is to define the theory on a finite size domain and then take the size of the domain to be infinitely. Let us consider the disc $\mathbb{D}_R$ with radius $R$ and impose the Dirichret condition at the boundary, that is, $\phi=0$ at the boundary. By integration by part, we can rewrite the action as follows.
\be
S[\phi] = - \frac{g}{8 \pi} \int_{\mathbb{D}_R} d^2 x \phi(x) (\Delta_x \phi)(x) \;.
\ee
With this Dirichret boundary condition, the Laplacian is invertible. The Green function is then given by
\be
G_R(z, w) = - \frac{1}{2 \pi} \log \left(\frac{R | z - w |}{| z \bar{w} - R^2 |} \right) \;.
\ee
In the large $R$ limit, we have
\be
G_R(z, w)\big|_{R \to \infty} = - \frac{1}{2 \pi} \log \left(\frac{| z - w |}{R} \right) \;.
\ee
Hence, the two-point function of a scalar field reads
\be
\left<\phi(z, \bar{z}) \phi(w, \bar{w}) \right>_{\mathbb{D}_R} =
\frac{4 \pi}{g} G_R(z, w) \;.
\ee
Since the generating function can be calculated exactly
\be
\left< e^{- \int d^d x J(x) \phi(x)} \right>_{\mathbb{D}_R} =
e^{- \frac{2 \pi}{g} \int \int d^d x d^d y J(x) G_R(x, y) J(y)} \;.
\ee
For simplicity, we take the normalization to be $g=1$. Then, the two-point function of a scalar field can be written by
\be
\left<\phi(z, \bar{z}) \phi(w, \bar{w}) \right>_{\mathbb{D}_R} = \log \left(\frac{R^2}{| z - w |^2} \right)
\ee
in the limit of large radius $R$.

The operator product expansion (OPE) is an important tool in CFT. For instance, let's take a set of local operators at nearby positions $\phi(x)$ and $\phi(y)$, and consider the product of those operators. This is written as the sum over all local operators.
\be
\phi_i(x) \phi_j(y) = \sum_{k} C_{ij}^{k}(x,y) \phi_k(y) \;.
\ee
The scale invariance can be used to determine the scaling behavior of the OPE coefficients $C_{ij}^{k}$ as $C_{ij}^{k}(x,y) \sim 1/|x-y|^{\Delta_i + \Delta_j - \Delta_k}$. Hence,
\be
\phi_i(x) \phi_j(y) = \sum_{k} \frac{c_{ij}^{k}}{(x-y)^{\Delta_i + \Delta_j - \Delta_k}} \phi_k(y) \;.
\ee

It has a great importance to explore the scale dependence of the effective field theory in understanding the physical phenomena, for instance, phase transition, universality, and critical phenomena. In order to describe those phenomena, the renormalization group is an essential tool in quantum field theory. Among several topics, let us focus on the effective field theory near the fixed points. In fact, the fixed points are described by the CFT. We make a perturbation to the action at the fixed points $S_*$ by some operators $\phi$.
\be
S = S_* + \sum_{i} g_i a^{\Delta_i - d} \int d^d x \phi_i(x) \;.
\ee

\end{document}